# Release Practices for Mobile Apps –
# What do Users and Developers Think?

Maleknaz Nayebi
SEDS lab, University of Calgary
Calgary, AB, Canada
mnayebi@ucalgary.ca

Bram Adams
MCIS lab, Polytechnique Montréal
Montreal, Canada
bram.adams@polymtl.ca

Guenther Ruhe
SEDS lab, University of Calgary
Calgary, AB, Canada
ruhe@ucalgary.ca

*Abstract*— **Large software organizations such as Facebook or Netflix, who otherwise make daily or even hourly releases of their web applications using continuous delivery, have had to invest heavily into a customized release strategy for their mobile apps, because the vetting process of app stores introduces lag and uncertainty into the release process. Amidst these large, resourceful organizations, it is unknown how the average mobile app developer organizes her app's releases, even though an incorrect strategy might bring a premature app update to the market that drives away customers towards the heavy market competition. To understand the common release strategies used for mobile apps, the rationale behind them and their perceived impact on users, we performed two surveys with users and developers. We found that half of the developers have a clear strategy for their mobile app releases, since especially the more experienced developers believe that it affects user feedback. We also found that users are aware of new app updates, yet only half of the surveyed users enables automatic updating of apps. While the release date and frequency is not a decisive factor to install an app, users prefer to install apps that were updated more recently and less frequently. Our study suggests that an app's release strategy is a factor that affects the ongoing success of mobile apps.**

## I. INTRODUCTION

Release planning of software products is a decision-centric problem that requires comprehensive information and knowledge. A release is a new or upgraded version of an evolving product that is characterized by a collection of new or modified features. Typical decisions made by release planners concern questions about the functionality (What to release?), time (When to release?) and quality (How good should the release be?) of an upcoming release. Depending on the market and risk involved, answering these questions requires thorough data collection and analysis about development and bug fixing progress, user escalations and competitors' progress. The recent practice of continuous delivery [1] is a parallel attempt to simplify release planning by releasing more frequently and reducing the scope of releases.

While substantial research has been done on release planning, on rapid releases and their impact on web and desktop applications [2, 3] (there is even a dedicated workshop on the topic of release engineering [4]), no such study has ever considered release planning for mobile apps. This is surprising, because Chuck Rossi (release engineering manager at Facebook) claims that "*mobile deployments are more challenging than Web deployments because we do not own the ecosystem, so we cannot do all the things that we would normally do*" [5]. Indeed, instead of releasing their app twice a day (like their Web site), Facebook and other major companies like Netflix release their mobile apps once every two weeks.

Since there are more than one million mobile apps across the major app stores, with thousands of developers specializing in mobile apps, competition is fierce [6]. Hence, release decisions matter even more for small, mobile app companies. In contrast to organizations like Facebook or Google, who have large budgets and substantial resources to develop release plans and tools, a typical mobile app company [7] consists of just a handful of developers, with 40% of all app developers having a separate main job and 21% working only part-time on apps. These small app vendors cannot afford to make mistakes and need to react in an agile way to market opportunities. In other words, making the right release decisions is key for them, without the luxury of having all required information or knowledge available or having ample time to process such information.

Because release practices as known in traditional software development might not apply in this new context, we are motivated to investigate the release practices that app developers follow. More specifically, we want to understand the extent to which release management of apps is based on developers' intuition compared to a formal rationale. In this paper, we refer to the thought processes and decisions going into release planning as *release strategy*. These also include the timing aspects of a release, such as decisions on the release date, the duration of a release cycle, and/or frequency of releases.

In particular, we performed one survey with mobile app developers and one survey with app users to get a better understanding of the perceived value and impact of the different release practices in use. We address the following research questions from a developer (RQ1-3) and a user (RQ4-6) point of view:

**RQ1:** What strategies do developers use to release apps?

*About half of the app developers has an explicit release strategy. Developers with a strategy tend to choose the strategy in the initial releases, then stick with it throughout the app's life.*

**RQ2:** What is the perceived impact of time-based release strategies on app development?

*The majority of developers are willing to bend their time-based strategies to accommodate users' feedback. Developers believe that apps with frequent updates likely deliver less changes of functionality and quality in each version.*

**RQ3:** What do developers think about the impact of release strategies on users?

*Developers believe that the rationale for release decisions affects user feedback. While the majority of developers believe that the time-based release strategies in general affect this feedback, the impact of release frequency is unclear.*

**RQ4:** To what extent are users aware of release updates and willing to update their apps?

*Almost all users are aware of mobile app updates, with only half of them always automatically updating their apps. For the majority of users, release date is not a deal-breaker, although they do prefer to install recently updated apps*

**RQ5:** What do users hesitate about for updating apps and what problems do they face after updating apps?

*Most participants have hesitated at one point to update apps and have experienced problems after updating their apps, especially due to lack of memory space, and phone or app crashes. When it comes to the actual problems caused by updates, device or app crashes are the major problem, followed by device slowdown, feature and functionality loss.*

**RQ6:** How do users perceive the value of frequent releases?

*Users have mixed feelings toward frequent app releases. Although they like apps with frequent updates and do not perceive it to include low quality apps, at the same time, frequent updates may be discouraging and could negatively affect users' decision for selecting, downloading or even uninstalling apps.*

In Section 2, we briefly highlight the current state of research on mobile apps in software engineering and we discuss the few empirical studies that exist in this domain. In Section 3, we describe our survey methodology and the conceptual model behind the study design. In Section 4, we characterize survey participants and introduce our classification of users and developers. We describe the results of the survey for different groups of users and developers in Section 5, then discuss our findings in Section 6. The limitations of our results are presented in Section 7, followed by the conclusions of the paper.

## II. RELATED WORK

Mobile apps are focused software products designed for mobile devices that are distributed through centralized mobile app stores [8]. The dynamism of these markets in combination with the multiple variables affecting market feedback (i.e., app rating, number of downloads and reviews) triggered substantial research in recent years. Considering app rating as an indicator of success for mobile apps, Linares-Vásquez et al. [9] analyzed the effect of API changes and fault tolerance of mobile apps on app success. Addressing the question of how effective apps have been created, reuse of mobile apps has been studied by Mojica et al. [10] and Linares-Vásquez et al. [11]. As another important quality or success characteristic, the security of apps has been analyzed [12, 13]. Only a small fraction of developers releases a large number of apps [14]. Those developers mainly belong to larger companies, while the majority of the developers work on their own.

The mobile app domain also has seen a number of survey studies to analyze app users' behavior [15-17]. Related to the software engineering aspects of mobile apps, the survey that is closest to our study is the one by Joorabchi et al. [18], who studied developers' challenges for cross-platform app development by performing a survey with 188 developers.

Although app stores have initiated a variety of research directions, the release engineering aspects of mobile apps have not been tackled so far, in contrast to web and software apps, where release management is well-established [19]. Several studies have been done to analyze the effect of rapid releases on software [1, 20, 21]. Khomh et al. [22] empirically analyzed the impact of the migration of the Mozilla Firefox release process towards shorter cycle time, on the software quality. They found that although bugs are fixed faster with rapid releases, proportionally fewer bugs are being fixed.

Recently, McIlroy et al. [23] empirically analyzed the update frequency of the top 10,713 mobile apps across 30 mobile app categories. Their results showed that 14% of the apps are updated frequently. 45% of the frequently-updated apps do not provide the users with any information about the rationale for the new updates. They also found that users highly rank frequently-updated apps and frequently-updated apps remain popular based on app stores' ranking. In contrast, we are interested in understanding why developers opt for a specific release strategy and whether there is in fact a plan, hence our study is not only limited to the frequency of releases. more, instead of deriving the impact of this strategy in terms of ratings and app store ranking, which might be affected by other factors, we directly include this as a question to the users.

In summary, while app stores are dynamic and many releases are observed in the app store [14], the how and why of practices and strategies for release management in this domain have not yet been studied.

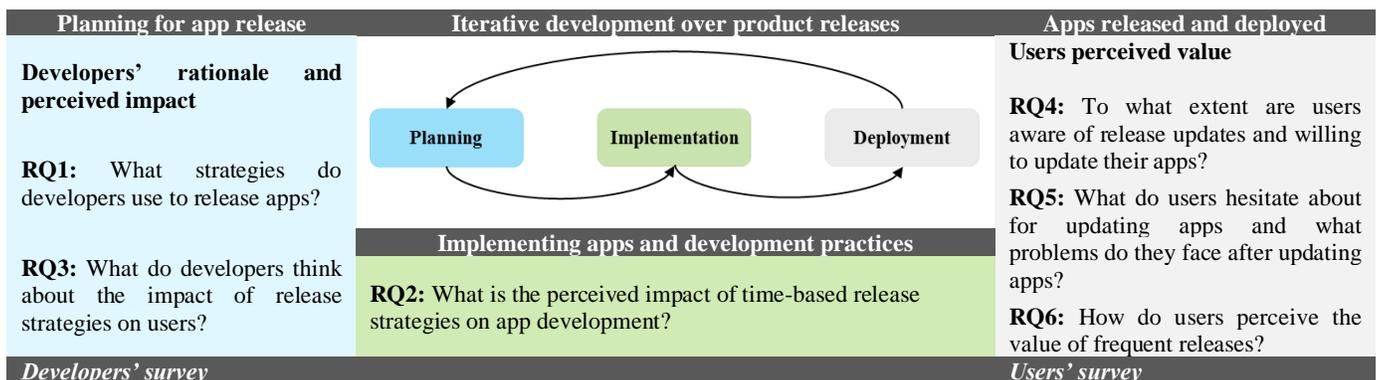

| Planning for app release | Iterative development over product releases | Apps released and deployed |
| --- | --- | --- |
| **Developers' rationale and perceived impact**<br><br>**RQ1:** What strategies do developers use to release apps?<br><br>**RQ3:** What do developers think about the impact of release strategies on users? | Planning → Implementation → Deployment<br><br>**Implementing apps and development practices**<br><br>**RQ2:** What is the perceived impact of time-based release strategies on app development? | **Users perceived value**<br><br>**RQ4:** To what extent are users aware of release updates and willing to update their apps?<br><br>**RQ5:** What do users hesitate about for updating apps and what problems do they face after updating apps?<br><br>**RQ6:** How do users perceive the value of frequent releases? |
| *Developers' survey* | | *Users' survey* |

**Figure 1. Structure of the study**

## III. METHODOLOGY

We used surveys to collect information from mobile app users and developers to describe and explain their behavior, attitude, and knowledge of using and releasing mobile apps. We performed two different survey studies, i.e., one with mobile app users and one with mobile app developers.

**Scope.** To study release strategies and practices, we targeted the planning, deployment and implementation phases of the app development lifecycle. The research questions related to each phase and the intended survey participant are shown in Figure 1. For a sequence of releases, we investigated the visibility of app updates for users, and investigated if users and developers think the frequency of releases affects the users' satisfaction and willingness to update apps. Also, we were interested to see what hesitations and actual problems are related to updating to a new mobile app version.

**Instrument.** We created a five-minute survey for mobile app users (including 16 close- and two open- ended questions) and a separate 10-minute survey for app developers (including 18 close- and six open- ended questions), both with descriptive design in order to explore and capture the description of release practices in mobile apps [24]. The surveys were approved by the University of Calgary Conjoint Faculties Research Ethics Board. To manage the survey distribution, we used the Qualtrics platform[1]. Both the Ethics board and seven grad and undergrad software engineering students evaluated the surveys in terms of understandability, reliability and validity of the questionnaire. Three of these participants are doing active research in release planning and engineering, and hence were able to perform content validity of the questionnaire, while the remaining four students performed face validity [25]. Their feedback was collected and the questionnaires were adjusted. The questionnaires can be found online[2].

**Participants.** For both surveys, we used a non-probabilistic convenience sampling method [26], which is a proper method to use when subjects are easily accessible. The prevalence of smart devices and mobile apps, especially for social networks, inspired us to advertise the survey on Twitter and Facebook.

For the *user survey*, our advertisements asked people to participate in a short academic survey on mobile apps and enter a raffle of five $15 amazon gift cards. We targeted the ad specifically at users of ages 15-85 in the US, Canada, Germany, and the UK who spoke English. We only published this advertisement on mobile news feeds. The ad reached 6,486 people and 2,197 clicks. In addition, we created a task in Amazon Mechanical Turk accessible only to people with experience with mobile apps. The advertisement and task were live for one week during August.

For the *developer survey*, we advertised the survey for one month on both Facebook and Twitter. We targeted the ad specifically at users in the US, Canada, Germany and UK of ages 15-60 who spoke English. We published this ad both in mobile and desktop news feeds. In total, we received 46 clicks on the ads. The advertisements were live for six weeks in August and September.

**Data Preparation and Analysis.** First, we excluded incomplete survey entries. We used descriptive statistics to characterize survey participants and to analyze survey results. In addition, we classified participants in each survey into different groups based on demographics (see Section 4). We then analyzed results for the different groups of participants. We used the Kruskal-Wallis test (a non-parametric test for analyzing significant differences between two or more groups) along with the Tamhane post-hoc test (conservative pairwise comparison) to analyze the significance of results between user groups and between developer groups.

For open questions, two software engineering students (including one of the authors) performed an open card sorting [27] technique using the Uxsort tool [28] to summarize the results. Card sorting is a set of activities for grouping and naming concepts or objects and to derive taxonomies from qualitative data [27]. In open card sorting, users define categories, while in closed card sorting the categories are pre-defined. The Cohen's Kappa [36] agreement between the two authors varied from 87% to 95%.

## IV. DEMOGRAPHIC ANALYSIS OF SURVEY PARTICIPANTS

Since we used a convenience sampling method [26] for both of the surveys, it is important to first get a better insight into the characteristics of the survey participants. For this, we classified both the surveyed users and developers based on demographic questions that we asked.

### A. Users

Through advertisements, we received 2,197 clicks and 674 people eventually participated in the survey. Among the answers, we excluded the users that left the survey incomplete and the ones who provided meaningless answers (like repetition of letters) to at least one open question. By excluding 20 entries, we analyzed the responses of 654 participants in the user survey.

The following information was gathered via the survey and is used to characterize the developers:

- The app stores the participant has ever download apps from.
- *Number of downloaded apps*: The number of times they personally installed apps on their devices.
- *Number of purchased apps*: The number of times they paid for an app.
- *Number of rated apps*: The number of times they rated apps.
- *Number of times wrote a review*: The number of times they have reviewed apps.

While the selection of the app stores was not exclusive (participants could select all the app stores they had experience working with), most of the participants (67.6%) used Google Play. After that, iTunes (with 26% of participants) and Microsoft Store (with 2.8% of participants)

---

**Table 1. Classification of users (a) and developers (b), applying classification algorithms for (c) usage, (d) success, (e) involvement and (f) experience.**

| (a)  Users | | |
|---|---|---|
| | *Low  usage* | *High usage* |
| *Low involvement* | *Category 'LILU'* 40.3% of participants | *Category 'LIHU'* 30.5% of participants |
| *High involvement* | *Category 'HILU'* 9.3% of participants | *Category 'HIHU'* 19.9% of participants |

| (b)  Developers | | |
|---|---|---|
| | *Low  success* | *High success* |
| *Low experience* | *Category 'LELS'* 5.6% of participants | *Category 'LEHS'* 36.1% of participants |
| *High experience* | *Category 'HELS'* 5.6% of participants | *Category 'HEHS'* 52.7% of participants |

| (c) Usage | | | |
|---|---|---|---|
| value | #downloaded apps (AQ1) | #purchased apps (AQ2) | Classification |
| 1 | ≤ 5 | = 0 | S = value(AQ1) + value(AQ2) If S < 5  Then *Usage is Low, Otherwise High* |
| 2 | 5 < # ≤ 10 | 0 < # ≤ 5 | |
| 3 | 10 < # ≤ 15 | 5 < # ≤  15 | |
| 4 | > 15 | >15 | |

| (d) Success | | | |
|---|---|---|---|
| value | #developed apps (AQ1) | #developers (AQ2) | Classification |
| 1 | 1 | ≤ 1 | S = value(AQ1)+ value(AQ2) If S < 5 Then *Success is Low, Otherwise High* |
| 2 | 1 < # ≤ 5 | 2 ≤ # ≤ 3 | |
| 3 | 6 < # ≤ 10 | 4 ≤ # ≤ 5 | |
| 4 | >10 | > 5 | |

| (e) Involvement | | | |
|---|---|---|---|
| value | #rated apps (AQ1) | #reviews (AQ2) | Classification |
| 1 | 1 | = 0 | S = value(AQ1) + value(AQ2) If S < 5  Then *Involvement is Low, Otherwise High* |
| 2 | 5 < # ≤ 10 | 0 < # ≤ 5 | |
| 3 | 10 < # ≤ 15 | 5 < # ≤  15 | |
| 4 | > 15 | >15 | |

| (f) Experience | | | |
|---|---|---|---|
| value | highest #downloads (AQ1) | best rating (AQ2) | Classification |
| 1 | <500 | ≤ 1 | S = value(AQ1)+ value(AQ2) If S < 5 Then *Experience is Low, Otherwise High* |
| 2 | 500 < # ≤ 50,000 | 2 ≤ # ≤ 3 | |
| 3 | 50,000 < # ≤ 500,000 | 4 ≤ # ≤ 5 | |
| 4 | >500,000 | > 5 | |

were used by our participants. Most of the users downloaded more than 15 apps. In terms of ratings, most of our participants rated apps five to ten times. Finally, most of the users wrote reviews less than 10 times and purchased an app less than five times.

Using the demographic information of participants, we classified the participants into four different usage groups based on their degree of usage and degree of involvement (in terms of providing feedback). We used the number of downloaded apps and the number of purchased apps as the *degree of usage* from app stores. Along with that, we used the number of rated apps and number of times they wrote a review as the *degree of involvement* for a participant. The groups were named *low involvement* and *low usage* (LILU), *low involvement* and *high usage* (LIHU), *high involvement* and *low usage* (HILU) and *high involvement* and *high usage* (HILU). Detailed descriptions of these categories and how we obtained them are presented in Table 1-(a). In all cases, the thresholds used are based on analysis of the survey results. We discuss them in more detail in Section 7.

As expected, people who rarely use app stores are rarely involved in feedback loops, hence the high involvement and low usage class, has the lowest number of participants. In reporting the survey results, we will compare the answers between these four user classes to analyze whether or not their perceived value of mobile app releases differs from one another.

*B. Developers*

Developers were persuaded to participate in the survey via advertisements, which received 46 clicks and 36 participants. All these 36 answers were complete and acceptable for our analysis. To analyze the characteristics of the participants, we asked questions about:

- *Number of developed apps:* The number of apps they developed personally or in a team.

- *Years of experience in app development:* The period of time during which they were involved in app development.
- *Highest number of downloads among apps:* The highest number of downloads among all the apps that they developed.
- *Rating of the most successful app:* Rating of their most downloaded app.
- *Years of experience in app development:* The period of time during which they were involved in app development.

We considered the number of developed apps, duration of app development and number of co-developers as the *developers' degree of expertise*. We used the highest number of downloads among apps and rating of the most successful app as the *degree of success* for the app developer.

Based on this data, we classified the developers into four groups, i.e., *low expertise* and *low success* (LELS, 5.6% of participants), *low expertise* and *high success* (LEHS, 36.1% of participants), *high expertise* and *low success* (HELS, 5.6% of participants), and *high expertise* and *high success* (HEHS, 52.7% of participants). The details of these classifications are presented in Table 1-(b). Based on common success criteria of apps, most participants had successful apps (88.8% of them) and they mainly have high experience in development (52.7% of them). This implies that our observations regarding release strategies correspond to strategies that worked in practice.

V.  FINDINGS

In this section, we summarize the findings from the developer and user surveys and discuss the differences and commonalities between the perception of developers and users regarding app releases. We do this by analyzing the differences between the four groups of developers and users identified in Table 1. If the p-value of Kruskal-Wallis was <0.05, we test for significant differences between each two groups by running Tamhane's test and we tagged those as ☑ Group1 vs. Group 2.

### A. Release Strategy and its Impact – Developers' Perception

Since mobile app developers have a wide variety of expertise and knowledge on software production in app stores [7], it is unclear to what extent they follow actual release management practices. In this section, we analyze these practices and to what extent their release strategies are based on rationale (explicit strategy) or intuition (no clear strategy).

#### 1) App Release Strategies (RQ1)

Release management practices were studied in medium to large software organizations and open source teams and the applicability of its principles into mobile app development is questionable. Mobile app teams tend to be smaller, with sometimes only one developer who needs to manage all of the development and release process. Moreover, app developers might be unexperienced in developing apps or developing software products in general [7] and not being experienced in the release management of software products.

**Half of the app developers (52.78%) follows a rationale-based release strategy.** To understand the state-of-the-practice of mobile app release management, we initially asked developers if they follow an established rationale for releasing their apps (release strategy) or whether they merely follow their intuition for defining what to release in the app version and when to release it.

☑ HEHS vs. LELS and LEHS (p-value=0.007 and 0.005 respectively): Our findings show that developers with high experience (and high success) tend to follow significantly more a rationale for releasing apps comparing to the ones with low experience in app development. The latter instead follow their intuition.

**The survey results show that release strategies are determined early in the app's lifecycle and most of the time developers do not change their strategy after releasing their apps.** Half of the participating developers with a certain release strategy stated that they followed the same strategy from the beginning of their apps' lifecycle. Also, the majority (66.67%) of participants that followed release strategies stated that they do not change it at all:

| Follow a release strategy | No 47.22% | Yes 52.78% | |
|---|---|---|---|
| Have changed release strategy | No 66.67% | Yes 33.33% | |
| How long been following a strategy | Same from start 50% | For half of the releases 31.25% | Only recently 18.75% |

**We identified six different categories of release strategies,** as summarized in Table 2. Among all the strategies, the time-based (80.0%) and marketing-based strategies (40.0%) are ranked first and second, respectively. Some app developers follow more than one release strategy. For example, the following developer considers both the *quality-driven* and the *time-based release* strategies:

*"Merging contributions, releasing alphas after each merge, freeze for beta with parallel alphas still going on, stable release when beta is good enough within one to six weeks."*

In this way, stabilizing a release within a reasonable time forms the rationale for the above developer. Another developer based his strategy on two pillars, i.e., offering localized (translated) versions of his apps to reach more markets and advertisement versions to entice users via popular forums or blogs. In this case, the release cycle time takes months instead of weeks.

> **Finding 1.** The majority of developers make rationale-based decisions for mobile app releases, which is observed significantly more often for developers with high experience.
>
> **Finding 2.** Time-based and marketing-based considerations are ranked as the first and second most popular rationale among developers, respectively.

#### 2) Perceived Impact of Time-based Release Strategy on Development (RQ2)

We also asked all developers with a release strategy (including those that never changed it) whether they could imagine changing their time-based strategy in order to achieve more success:

| | Disagree | Neutral | Agree |
|---|---|---|---|
| Change time-based strategy to achieve higher success | 22.23% | 13.77% | *64%* |

**In general, the main drivers for the developers to**

**Table 2. Release strategies as stated by developers.**

| Category | Description | Frequency of categories mentioned by developers | Total |
|---|---|---|---|
| Time-based | Weekly or bi-weekly | 45% | 80.0% |
| | Yearly | 15% | |
| | Quantitative descriptions (agile releases, releasing often, …) | 10% | |
| | Time-based strategy for pre- or post-release versions | 10% | |
| Marketing considerations | App localization and country targeting | 25% | 40.0% |
| | Blog spill-outs | 10% | |
| | Promotions | 5% | |
| Quality (test) driven | Test coverage and internal quality assurance | 10% | 25.0% |
| | Over the air installation and testing | 10% | |
| | Test with a portion of customers | 5% | |
| Feature-based | Implementing planned features | 10% | 20.0% |
| | Release each implemented feature | 10% | |
| Size-based | Size of the update | 10% | 10% |
| Occasional | Release app during the launch of a new device of a well-known brand | 5% | 5% |

**change their time-based release strategy are user feedback and personal experience with earlier releases**. In particular, the participants who agreed to change their time-based strategy, stated the following reasons for doing so:

| | |
|---|---|
| Accommodating user feedback | 36.1% |
| Gained experience in similar cases | 16.6% |
| Do not feel obligated to follow self-defined schedule | 13.9% |
| Occurrence of specific occasions or events | 8.3% |
| Status of competitor apps and their releases | 5. 6% |
| Gaining higher market visibility by analyzing country targeting states | 8.1% |
| Following the main schedule in addition to hot fixes | 6.3% |
| Time needed to stabilize app due to the results of over-the-air testing platforms such as Testflight or Hockeyapp | 5.1% |

On the other side, the disagreeing participants stated the following reasons for not changing their release strategy:

| | |
|---|---|
| Users are waiting for the release | 35.7% |
| Special occasion may pass | 27.1% |
| Competitor apps would release the product sooner | 24.6% |
| Limited resources and capacity | 12.6% |

**The majority (61.1%) of developers believe that apps with frequent updates (more than once per 3 weeks) deliver less functionality and quality changes in each version**. On the other hand, developers were unsure whether apps with infrequent updates had an overall higher quality:

| | | | |
|---|---|---|---|
| Apps with frequent updates deliver less changes in each version | Disagree 27.8% | Neutral 11.1% | Agree 61.1% |
| Less releases means higher quality | Disagree 38.9% | Neutral 22.2% | Agree 38.9% |

☑ HEHS vs. LEHS (p-value= 0.003): Developers with high experience (and high success) believe significantly more that apps with frequent updates have lower quality than developers with low experience.

Furthermore, **a minority of the participants stated that the apps' time-based release strategy may affect the development practices, team size, and effort needed to develop them**. Whereas 44.5% of the participants did not agree with this statement, 36.12% state that their selected time-based release strategy changed the way they developed the app in terms of number of developers involved, test coverage, code quality, code review and so on. In particular, they stated the following changes:

- More development effort.

  *"...more developers and planning and team meetings needed".*

  *"...somehow, more work!"*

- Less provided functionality than targeted.

  *"...Spend time on testing, (so) we stop (including) new feature developments*

- Change in packaging and testing the apps.

  *"... I pack all changes each 2 weeks and give around 4 weeks to stabilize*

  *"If you're looking for a 'Hollywood-blockbuster' style (where v1.0 has a big download spike, then quickly trails off - typical for many apps), you'll want a long beta test with a healthy amount of testing and development before v1.0, If you're looking for a 'slow*

*growth' style (particularly common f the app is for an already-well-known company [...]) you should still beta test, but it's OK to launch with more of a 'minimum viable product' and grow from there."*

- Change in design and implementation of the app.

  *"... We need more design for features in each release so we need to pay more for designers and hire one part time."*

  *"I need to test and develop and design overall a lot of cathartic work so I do a bit of each instead of finishing one and go to the other."*

Despite the majority of developers claiming no effects on development practices, team size and development, 44.5% of participants agree that frequent releases need more developers:

| | Disagree | Neutral | Agree |
|---|---|---|---|
| Time-based release strategy affects development practices, team size and development | *44.5%* | 19.5% | 36.1% |
| More frequent updates need bigger teams and more effort | 36.1% | 19.5% | *44.5%* |

The contradictions between some of these results suggest the need for further research on this topic.

> **Finding 3.** Developers are flexible to deviate from their apps' time-based release strategies, mainly to gain better user feedback.
>
> **Finding 4.** Developers believe that apps with frequent updates deliver less changes and do not need additional development effort.

### 3) Perceived Impact of Release Strategy on Users (RQ3)

Prior research has shown that the feedback provided by user reviews, emergence of successful apps and the general ecosystem dynamics affect new versions of apps [29-31]. In our survey as well, **44.5% of developers agree that the release strategy affects user feedback** in the form of ratings, reviews, and number of downloads:

| | Disagree | Neutral | Agree |
|---|---|---|---|
| Release strategies affect user feedback | 36.1% | 19.4% | *44.5%* |

☑ HEHS vs. LEHS (P-value: 0.029): Developers with high experience agreed significantly stronger compared to the developers with low experience (and high success).

In particular, the majority of participants (61.1%) agreed that time-based strategies affect user feedback (rating, number of downloads, and reviews), yet developers are unsure about the exact effect of update frequency on app users. While 38.9% of participants agreed that frequent app releases have a negative impact on user feedback, a similar number disagreed.

| | Disagree | Neutral | Agree |
|---|---|---|---|
| Frequent updates causes negative user feedback | 38. 9% | 22.2% | 38.9% |
| Time-based release strategy affects users' feedback | 27.8% | 11.1% | *61.1%* |

> **Finding 5.** The release strategy affects user feedback, something that developers with high experience are most convinced about.
>
> **Finding 6.** Developers believe that, in general, the release strategy affects user feedback.

*B. Release Strategies and its Impact – Users' Perception*

Having learned the developers' perspective on release strategies, we now investigate the users' perception.

*1) Awareness of app Updates (RQ4)*

Many mobile apps provide an automatic updating facility that automatically updates an app to its latest release, without user intervention. Apart from making life easier for users, such a functionality helps a company focus on maintaining just one app version in the field as opposed to having to maintain a heterogeneous mixture of old releases. Whereas for desktop systems like browsers large companies have a clear idea about which versions are still being used, for mobile app companies, especially the smaller ones, this is not necessarily the case.

To investigate the app users' stance towards automatic updates, we asked users explicitly about their awareness of updates to their installed mobile apps. The results showed that **almost all users (96.1%) notice the app updates and are aware of it**. Also, users do not blindly enable automatic updates for all mobile apps, as **41% of the users only partially and 14.1% never enable automatic updates.**

| Noticed app updates | Yes 96.1% | | No 3.9% | |
|---|---|---|---|---|
| Allow automatic updating of apps | Always *44.9%* | Partially 41% | | Never 14.1% |
| | | Rarely 18.5% | Monthly *45.6%* | Weekly 25.5% More often 10.4% |

Of the users that do not always allow automatic updates, 18.5% rarely update their mobile apps and 45.6% update less than once a month. In particular, power users seem pickier regarding automatic updates.

☑ LUHI vs. LULI (P-value=0.002): Users with low usage and high involvement allow automatic updates significantly less often than users with low usage and low involvement.

☑ HUHI vs. LULI (P-value = 0.007): Also, users with high usage and high involvement allow automatic updates significantly less often than users with low usage and low involvement.

Finally, the majority of users (59.6%) state that the recency of the latest version of an app has never prompted them to buy/install an app. However, 61.9% state that, given the choice between two equivalent apps, they would choose the one having more recent release date. So, while the last release date could potentially break the tie between two similar apps, this information in app stores is not decisive for choosing an app for 40.38% of the users. Power users again exploit this kind of information more than others do:

| | No | Yes |
|---|---|---|
| Prefer to install app with recent release date in case of equal functionality and quality | 38.1% | *61.9%* |
| Ever decided to install an app because of last release date | No *59.6%* | Yes 40.4% |

☑ HUHI vs. LULI (p-value: 0.025): We found that users with high usage and involvement consider release date to be significantly more important compared to the users with low usage and low involvement.

Given that the "last release date" is available along with other app information in the app stores, 79.2% of participants reported that they observed regularity in release cycles (35.4% did this rarely, while 43.8% did this frequently). Among the participants who observed regularities, 28.7% reported having observed weekly updates and 47.4% reported having seen monthly updates. 2.4% of users reported other types of regularities such as *bi-weekly updates*, *updates by closing the app*, *restarting the device or logging-out* of the application:

| | Never 22.1% | Rarely 34.1% | | Occasionally *43.8%* | |
|---|---|---|---|---|---|
| Observed regularity in app updates | | Occasional updates 16.5% | Yearly updates 5.1% | Monthly updates *47.4%* | Weekly updates 28.6% Other 2.4% |

Further, we did not find a correlation between the manual update habits and the observed regularity in app updates. This means that for example, the users who usually update their apps once a month manually, did not necessarily report monthly regularity in app updates.

**Table 3. Reasons of users to hesitate to update apps.**

| Category | Type of concern | Frequency of statement | Total |
|---|---|---|---|
| *Quality of the app* | Device and app crash | 9.0% | *37.40%* |
| | Security and privacy | 8.4% | |
| | Bugs and compatibility concern | 13.8% | |
| | Speed | 4.7% | |
| | Data loss | 1.5% | |
| *Release and version specification* | Download and update size | 5.6% | *16.40%* |
| | Lack of update description | 2.0% | |
| | Time to stabilize before nload | 1.2% | |
| | Improper timing for update | 5.0% | |
| | Frequent updates | 2.6% | |
| *Device-related issues* | Space | 13.3% | *15.60%* |
| | Battery problems | 2.3% | |
| *Functionality of the app* | Uninteresting update | 5.2% | *12.70%* |
| | Feature and functionality loss | 7.5% | |
| *Negative arguments* | Negative experience from past updates of the app | 1.4% | *10.10%* |
| | Reviews | 5.2% | |
| | Anecdotal reports | 3.5% | |
| *App-related issues* | Unused app | 4.2% | *5.30%* |
| | Advertisement and paywall | 1.1% | |
| *Gut feeling* | -- | -- | *2.50%* |

**Table 4. Actual problems that users faced after updating mobile apps.**

| Category | Type of concern | Frequency of statement | Total |
|----------|-----------------|------------------------|-------|
| *Quality of the app* | Phone and app crash | 51.2% | *81.60%* |
| | Speed | 13.2% | |
| | Bugs | 9.1% | |
| | Data loss | 3.1% | |
| | Compatibility concern | 2.8% | |
| | Security and Privacy | 2.2% | |
| *Functionality of the app* | Feature and functionality loss | 10.3% | *11.60%* |
| | Undesirable UI | 1.3% | |
| *Device-related issues* | Space | 2.5% | *4.50%* |
| | Battery draining | 2.0% | |
| *App-related issues* | Advertisement and Paywall | 2.3% | *2.30%* |

> **Finding 7.** Users are aware of mobile app updates. Less than half of them always turns on automatic updates, especially users with low usage and low involvement.
>
> **Finding 8.** While the release date is not a decisive factor to install an app, users prefer to have apps that were updated more recently.
>
> **Finding 9.** Users observed regularity in the release cycle of mobile apps independent from their preference for manual or automatic updating of apps.

### 2) Users' Hesitations with App Updates (RQ5)

**60.3% of participants have some hesitations to update mobile apps, and almost half of the participants (47.7%) have encountered a problem after updating apps**.

| Have hesitation for updating apps | No | Yes |
|---|---|---|
| | *39.7%* | *60.3%* |

| Encountered problems by updating apps | No | Yes |
|---|---|---|
| | *52.3%* | *47.7%* |

We categorized the participants' hesitations into seven categories in Table 3, while the actual problems experienced after app updates are categorized in four groups in Table 4.

**The top reason why users hesitate to update mobile apps (stated by 13.3%) is the lack of memory space on their device**. Phone and app crashes (9% of participants), security and privacy (8.4% of participants), and feature and functionality (7.5% of participants) are the next most common reasons:

*"... (App updates) sometimes takes too long and jacks up my phone." [Speed]*

*"... There were reviews how buggy it was." [Reviews - Bugs]*
*"Didn't want to lose date/game scores." [Data loss]*

*"Requesting permissions for things the app doesn't need like contacts info and pictures." [Security and Privacy]*

*"Occasionally an update will cause configuration and compatibility problems." [Compatibility concern]*

When it comes to actual problems experienced, **51.2% of the users reported device or app crashes as the major problem they faced after updating their apps**. After that, low speed of the app (13.2% of participants), feature and functionality loss (10.3% of participants) and bugs (9.1% of

participants) were reported as the largest problems users faced:

*"(App) starts sending me messages without asking me and I think it might have started accessing my data without my permission." [Security and privacy]*

*"Incompatible features, higher battery usage, memory usage, ads." [Compatibility concern, Battery draining, space, advertisement]*

*"Usually the app would crash or other problems so I try to update as little as possible. Unless there are new features." [Device and app crashes]*

☑ HUHI, HULI vs LUHI, LULI (P-value: 0.009): The users with low usage (both in LUHI and LULI) encountered significantly more problems after updating their apps, compared to those who make more use of the apps (both in HULI and HUHI categories).

> **Finding 10.** The majority of users have concerns regarding updating apps and have faced update-related problems before. Users with low usage encountered significantly more problems comparing to users with high usage. Device and app crashes were the main reported problems.

### 3) Awareness about Release Frequency (RQ6)

**Overall, users have mixed feelings toward the frequency of releases, so while they like it at times they may also find it bothersome.** 70.8% of users agreed that they are happy with their installed apps that have frequent updates. At the same time, 35.3% users claimed that they did uninstall apps because of too frequent releases:

| Apps with frequent update | Undesirable | Neutral | Desirable |
|---|---|---|---|
| | 18.5% | 10.7% | *70.8%* |

| Uninstall apps because of frequent update | No | Yes |
|---|---|---|
| | 64.7% | 35.3% |

Also, among apps with the same functionality and quality, 54.6% of participants prefer to install apps with less need for frequent updates. On the other side, frequent updates do not imply lower quality to the users. 66.2% of participants disagreed that frequently updated apps have lower quality.

| Prefer to Install apps with less frequent updates in case of equal functionality and quality | No | Yes | |
|---|---|---|---|
| | 45.4% | *54.6%* | |

| Apps with frequent update have lower quality | Disagree | Neutral | Agree |
|---|---|---|---|
| | *66.2%* | 11.5% | 22.3% |

☑ HULI vs. LUHI and HUHI (P-value=0.000 and 0.0002 respectively): The users with high usage and low involvement uninstall apps significantly more than the users with high involvement (both LUHI and HUHI).

☑ HULI vs HUHI (P-value=0.005): The users with high usage and low involvement prefer to download apps with frequent updates significantly less than the users with high usage and high involvement.

> **Finding 11.** Users have mixed feelings toward frequent app releases. They like apps with frequent updates but at the same time, frequent updates may be discouraging and could negatively affect users' decision.

## VI. DISCUSSION

During our analysis of the update attitude of mobile app users, we found that manual updating of mobile apps is still quite common (RQ4), despite the automatic update facilities of mobile app platforms. This might be linked to the sometimes negative experience of users regarding the frequency of releases (RQ6). Moreover, this might result in a huge diversity of running versions of an app among different users and cause maintenance issues [32]. From this point of view, we asked both users and developers about their perception of frequent app updates.

Developers believe that the time-based release strategy in general affects user feedback, but the effect of release frequency in particular is unclear (Finding 6). On the other side, users have mixed feelings about release frequency: while they like apps with frequent updates, they might find it discouraging or might even uninstall the app for this reason (Finding 8). The perception of developers and users on the impact of release frequency is demonstrated in Figure 2 – (a) and Figure 2 – (b).

> **Challenge 1.** The impact of release frequency on mobile app users is unclear.

The results of our analysis of the relation between release frequency and app quality was ambiguous. Half of the developers believed that less releases meant higher quality, while on the users' side the majority believed that less releases does *not* necessarily imply better quality of the app. The feelings of developers and users about the relation between release frequency and app quality is demonstrated in Figure 2 – (c) and Figure 2 – (d).

> **Challenge 2.** The relation between release frequency and app quality is unclear.

There is an ongoing discussion in the software engineering community on the commonalities and differences of mobile apps and traditional software products. The results of this survey show that the release management of mobile apps is different in a variety of aspects from traditional software release management. In particular, the problem of when to release and release readiness [19] are affected by other variables. For example, the frequency of releases, reluctance of updating apps on one side in conjunction with the development effort needed affect the decision of when to release an app.

In particular, developers who participated in our survey mostly believed that the choice of a frequent release strategy does not affect development process. However, they did believe that a larger team might be necessary to accomplish this. Given that mobile apps traditionally have been developed by smaller organizations [6], the decision to support frequent releases might have a large impact on the average mobile app company compared to release practice for traditional software development. Empirical investigation of this phenomenon is needed.

> **Challenge 3.** While the developers believe that release strategy and frequency do not affect development practices and total effort needed, empirical analysis using app store mining will be necessary to holistically evaluate and compare to traditional software release practices.

The qualitative results of the developer survey provide insight into how common release strategies are (only half of the developers had an explicit strategy) and the specific release strategies that they follow such as size-based strategies. While these strategies are not known for releasing in traditional software development and as the size of updates is a concern for users (see table 3), the popularity and success of such strategies needs further research.

> **Challenge 4.** The popularity and efficiency of release strategies and comparison with release strategies in traditional software products is needed to define the state of practice in mobile apps' release management.

A recent survey on challenges in product management revealed that half of the product managers consider "the process" the biggest challenge [33]. The survey included almost 900 respondents from around the globe and from a wide variety of industries and company sizes. While this is true for "traditional" software-based products, in this paper we have investigated some specific questions related to processes in releasing mobile apps. Further research is necessary to answer questions such as

*Which release strategies could provide higher popularity and visibility for an app?*

*What decision variables should be considered by developers for releasing their app?*

Inspired by app stores and their emerging impact on software engineering, it was predicted that requirement engineering will hugely be affected by users and their feedback [35]. This study could be useful for developers and app owners to understand and address users' problems and hesitations. For example, *better app descriptions* could help users better understand changes, explicitly defining *consumption of device resources* could address battery draining concerns [18] and *quality assurance and usability testing could address* the huge concerns regarding quality issues of new app versions [18, 34,37].

## VII. THREATS TO VALIDITY

Several limitations should be considered while interpreting and reusing the reported results.

**Construct validity.** Although we experimented with automated clustering of the users and developers, which led to seemingly random groups without clear rationale, we used manual clustering based on clear criteria. We believe that users with different degree of usage of mobile apps and involvement in providing feedback experience different feelings about the release practices, while for developers the role of experience in managing and valuing the releases is undeniable. However, one should consider the following as the limitations of these classification:

- The classification resulted in unbalanced grouping, especially in terms of developer success. This is partly by design, since the thresholds that we used to cluster apps are based on common success criteria for apps (divide it into upper and lower half), and we want to learn the release strategies used by successful app developers. Furthermore, despite our survey setup, convenience sampling has the inherent risk of attracting only particular subpopulations.

- The terms 'low' and 'high' for the degree of usage and involvement of users, and for the degree of experience and success of developers are relative. So we categorized a user who downloaded more than 15 apps and never purchased an app as a user with low usage relative to the other users in the survey.

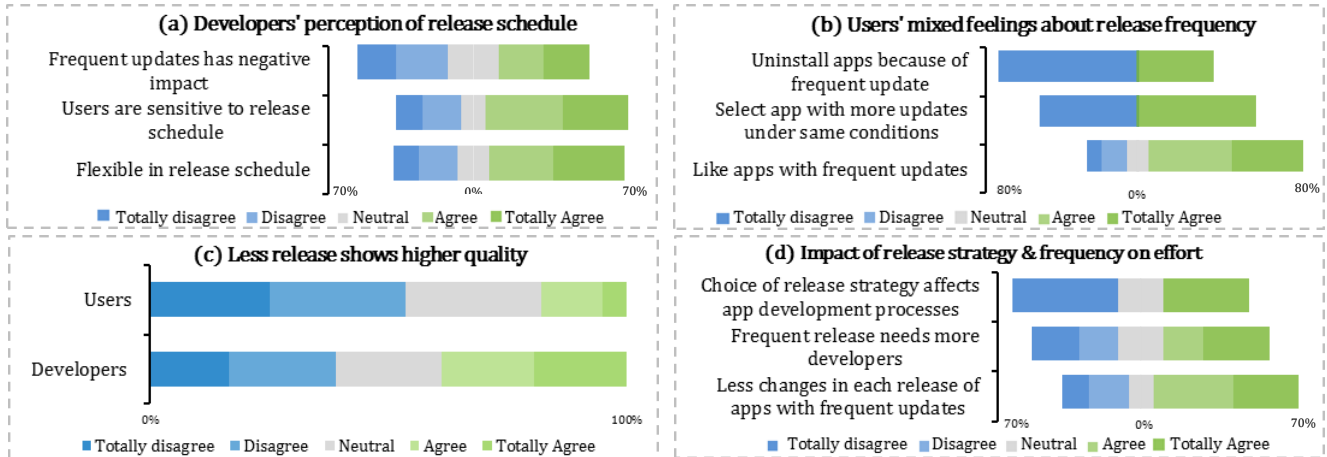

**Figure 2. Summary of findings that led to challenges.**

In addition, the sample size of users is much larger than that of developers (674 in comparison to 36). This unbalanced sample might endanger the validity of the study. However, the number of app users is much larger than the number of mobile app users as well.

Also, in the surveys we only targeted the planning, implementation and deployment phases of the software development lifecycle as they likely have the most effect on release management. We designed questions based on our own release planning and engineering knowledge and from the available literature about software projects. Many other release management practices and topics could be addressed, but were not in order to keep the survey focused (to avoid scaring participants). We have mitigated this risk by designing open-ended questions.

**Internal validity.** The survey results reflect the subjective opinion of the participants, which may be different from reality. For example, when comparing quality and functionality offered in each release (RQ2) or talking about battery drain (RQ5), the meaning of "high" and "low" are subjective.

**External validity.** App stores attract many developers and users from all around the world. Compared to the total number of involved users and developers, our sample in this study includes a limited number of users (645) and developers (36). Yet, this is still a larger sample than previous studies on user behavior [15-17], and, given the large effort going into the planning and analysis of a survey, in general the study of developers in this context is rare [18]. Also, we used convenience sampling [26] to obtain responses for both surveys. This increases the risk of bias of targeted population. For the user's survey, we mitigated this risk by targeting the advertisement only to Facebook and Twitter app users to make sure that they are familiar with the usage of mobile apps. Finally, the open questions and the qualitative findings, even if the sample would be biased, provide a wealth of information for further research.

## VIII. CONCLUSIONS

We performed two surveys (i) with 36 app developers to extract release strategies for mobile apps in general and the time-based release strategies in particular, to find the developers' rationale for release decisions, and the impact of their choice of strategy on development process and users; and (ii) with 654 app users to explore problems they have with updating apps and their feeling toward strategies with frequent updates. Our findings show a partial difference between release strategies of mobile apps and traditional software. In particular, we found relatively unknown marketing-based and size-based strategies in this context. Our study showed that half of the developers follow rationale-based release strategies and that time-based strategies are used by the majority of these developers. The mixed feelings of users about the frequency of releases and the uncertainty of developers about the impact of updates should be further supported by complementary app store analysis. Moreover, our study shows that release decisions should be re-evaluated and aligned toward user's needs and convenience.